\documentclass[preprint,12pt]{elsarticle}

\usepackage{amssymb}

\usepackage{color}

\journal{Nuclear Physics A}

\begin{document}

\begin{frontmatter}



\title{Neutrino-induced reactions on $^{18}$O and implications of $^{18}$O mixture in water Cherenkov detectors on supernova neutrino events}


\author{Toshio \textsc{Suzuki$^{a,b,c}$}\footnote{E-mail: suzuki.toshio@nihon-u.ac.jp}}
\author{Ken'ichiro \textsc{Nakazato$^{d}$}}
\author{Makoto \textsc{Sakuda$^{e}$}}

\address[]{Department of Physics, College of Humanities and Sciences, Nihon University, Setagaya-ku, Tokyo 156-8550, Japan}
\address[]{NAT Research Center,3129-45 Hibara Muramatsu, Tokai-mura, Naka-gun, Ibaraki 319-1112, Japan}
\address[]{School of Physics, Beihang University, 37 Xueyuan Road, Haidian-qu, Beijing 100083, People’s Republic of China}
\address[]{Faculty of Arts and Science, Kyushu University, Fukuoka 819-0395, Japan}
\address[]{Physics Department, Okayama University, Okayama 700-8530, Japan}



\begin{abstract}
Neutrino-nucleus reaction cross sections on $^{18}$O are evaluated by shell-model calculations and compared with those on $^{16}$O.
Important contributions from Gamow-Teller transitions are noticed for $^{18}$O ($\nu_{e}$, e$^{-}$) $^{18}$F in contrary to the case for $^{16}$O, where spin-dipole transitions are dominant contributions.
Calculated cross sections for $^{18}$O ($\nu_e$, e$^{-}$) $^{18}$F are shown to be larger than for $^{16}$O at low neutrino energies below 20 MeV in natural water with the 0.205$\%$ admixture of $^{18}$O due to the lower threshold energy (1.66 MeV) for $^{18}$O than that for $^{16}$O (15.42 MeV). 
The resulting electron spectra, that is, the cross sections as functions of emitted electron energy $T_e$,  
are also shown to be quite different, reflecting the different threshold energies. 
The electron spectra from ($\nu_e$, e$^{-}$) reactions on $^{18}$O and $^{16}$O in water Cherenkov detectors for supernova neutrino detection are investigated for both the cases with and without the neutrino oscillation and compared with those of the neutrino-electron scattering. 
It has been shown that the contribution from $^{18}$O (0.205$\%$ mixture) enhances the rates from $^{16}$O by 60$\%$ for the case without the oscillation and by 20-30$\%$ for the case with the oscillation below $T_e$ =20 MeV.
For the case with the neutrino oscillation, the event rates for $^{18}$O and $^{16}$O become comparable to those of the neutrino-electron scattering.
However, their rates at low energy ($T_e <$ 20 MeV) are much smaller than those of the neutrino-electron scattering, which is important for the pointing accuracy to the supernova direction. 
\end{abstract}




\end{frontmatter}


\section{Introduction}
\label{s1}
Water Cherenkov detectors are powerful tools to probe supernova neutrinos and study their properties.
Superkamiokande \cite{superK} is under way and Hyperkamiokande \cite{hyperK} is planned 
to determine the neutrino mass hierarchy and CP-violating phase as well as supernova explosion dynamics \cite{Langanke}.
Both charged-current and neutral-current neutrino-nucleus reactions on $^{16}$O have been studied by shell model \cite{SC2018} and CRPA calculations \cite{KL}. 
Dominant contributions to the cross sections come from spin-dipole (SD) transitions.
The SD strengths, charged- and neutral-current total and partial reaction cross sections for various particle and $\gamma$ emission channels have been evaluated with the Hauser-Feshbach statistical model \cite{SC2018,KL}. 

Here, we study $\nu$-induced reactions on $^{18}$O, which has isotope abundance of 0.205$\%$. 
Gamow-Teller (GT) transitions give considerable contributions to the charged-current reaction cross sections for $^{18}$O ($\nu_{e}$, e$^{-}$) $^{18}$F.
Experimental data for the GT strength in $^{18}$O was obtained by ($^{3}$He, $t$) reactions on $^{18}$O \cite{ox18gt}.
Cross sections for $^{18}$O ($\nu_{e}$, e$^{-}$) $^{18}$F are evaluated with the use of an effective axial-vector coupling constant $g_A^{eff}$ determined from the experimental GT strength.

Charged-current reaction $^{18}$O ($\nu_e$, e$^{-}$) $^{18}$F caused by the admixture of $^{18}$O in natural water was calculated previously and pointed out to account for about 10$\%$ of electron events  
induced by solar neutrinos generated by $^{8}$B $\beta$ decay \cite{Haxton}.
Elastic $\nu$-$e^{-}$ scattering is the main source of the electron events.
Taking account of the isotopic abundance of $^{18}$O, sum of $\nu$-induced reaction cross sections on $^{16}$O and $^{18}$O were evaluated for supernova spectra, which were taken to be Fermi distributions with temperatures $T$ = 3-10 MeV \cite{Haxton}.
The temperature of $\bar{\nu}_e$ was suggested to be $T_{\bar{\nu}_e}$ =4-5 MeV from the measurement of SN1987A neutrinos at Kamioka \cite{Kamioka} and IMB \cite{IMB}, but no observational information was available for the temperatures of $\nu_e$ and $\nu_x$ where $x$ = $\mu$, $\tau$ or $\bar{\mu}$, $\bar{\tau}$. 
There were fairly large uncertainties in the supernova neutrino spectra.
Supernova model calculations lead to a hierarchy for the temperatures, $T_{\nu_e} \leq$ $T_{\bar{\nu}_e}<$ $T_{\nu_x}$, where $T_{\nu_x}$ was predicted to be as high as 8 MeV \cite{Wilson}.
Nucleosynthesis of elements produced by $\nu$-processes was studied with the use of temperatures that satisfy this hierarchy;
($T_{\nu_e}$, $T_{\bar{\nu}_e}$, $T_{\nu_x}$) = (3.2, 5, 8) MeV \cite{Woosley} or (4, 4, 8) MeV \cite{WW95}.
A lower temperature for $T_{\nu_x}\approx$ 6 MeV was pointed out to be favored from
constraints on the abundance of $^{11}$B 
obtained by $\nu$-process and galactic chemical evolution \cite{Yoshida2005,Yoshida2006}. 
The observed solar-system abundances of $\nu$-process elements, $^{138}$La and $^{180}$Ta, are found to be consistently reproduced by taking $T_{\nu_e}$ $\approx$ $T_{\bar{\nu}_e}$ = 4 MeV \cite{Hayakawa1,Hayakawa2}. 
While temperatures of the Fermi distributions have been updated, another analytical form for the neutrino spectra  
called modified Maxwell-Boltzmann distribution 
was proposed \cite{KRJ} and became more commonly used than the Fermi distributions.
The modified Maxwell-Boltzmann distribution has two characteristic parameters, which are average energy and spectral pinching.
Recent supernova models lead to spectra with smaller average energy for $\nu_x$, that is, $T_{\nu_x}$ $\approx$ $T_{\bar{\nu}_e}$, but with large high energy components produced in the accretion phase of the supernova explosions \cite{Naka2013,Siverding1,Siverding2}.  
Here, possible effects of the $^{18}$O mixture on the count rate of supernova $\nu$ events in water Cherenkov detectors are examined with the use of recent realistic neutrino spectra.
Effects of neutrino oscillations,  
which exchange $\nu_e$ and $\nu_x$, on the count rate are also investigated.

In Sect. 2, the GT strength in $^{18}$O is obtained by shell-model calculations, and compared with the experimental data.
Then, $\nu$-induced reaction cross sections for $^{18}$O are evaluated for both charged- and neutral-current channels, and
compared with those for $^{16}$O.
Event spectra for emitted electrons induced by reactions on natural water are also examined.
In Sect. 3, contributions of $^{18}$O mixture to the count rate of supernova $\nu$ events in water Cherenkov detectors are estimated. 
Summary is given in Sect. 4.

\section{$\nu$-induced reactions on $^{18}$O}
\label{sec2}
\subsection{Gamow-Teller strength in $^{18}$O}
We first evaluate GT strength in $^{18}$O by shell-model calculations with the use of SFO-tls Hamiltonian \cite{SFO} in $p$-$sd$ shell.
The Hamiltonian, SFO-tls, was used to obtain $\nu$-induced reaction cross sections in $^{16}$O \cite{SC2018}.
The $B(GT_{\pm})$ is defined as
\begin{equation}
B(GT_{\pm}) = \frac{1}{2J_i +1} |<f|| q \sum_{k} \vec{\sigma}_{k} t_{\pm}^{k} ||i>|^2
\end{equation}
where $J_i$ is the spin of the initial state, $t_{-}|n>$ =$|p>$, $t_{+}|p>$=$|n>$ and 
$q$ is the quenching factor for the axial-vector coupling constant, $q$ =$g_A^{eff}$/$g_A$.
The sum runs over all nucleons.  
The quenching factor is determined to reproduce the experimental sum of the strength, $S$ = 4.06, measured up to the excitation energy $E_x$ = 12 MeV.
It is obtained to be $q$= 0.88.
Calculated $B(GT_{-})$ and the experimental data \cite{ox18gt} are shown in Fig. 1.

\begin{figure}[htbp]
\vspace*{-0.5cm}
\begin{minipage}{0.45\hsize}
\begin{center}
\hspace*{1cm}
\includegraphics[scale=0.55]{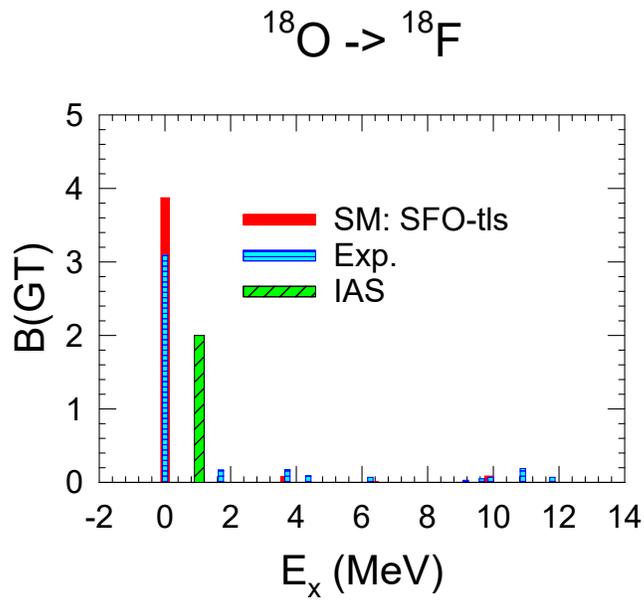}
\end{center}
\end{minipage}
\vspace*{-3.5cm}
\caption{\small 
Gamow-Teller strengths obtained by shell-model calculations with the use of SFO-tls and experimental data \cite{ox18gt} are shown by red solid and blue hatched histograms, respectively. The quenching factor for $g_A$ is taken to be $q$ =0.88.
Green dashed histogram denotes Fermi contributions, $B(F)$, from the transition to the isobaric analog state (IAS), $^{18}$F (0$^{+}$, 1.04 MeV, T=1).} 
\label{o18gt}
\end{figure}

The GT transitions from the ground state of $^{18}$O (0$^{+}$, T=1) to the 1$^{+}$ states in $^{18}$F with isospin T = 0, 1 and 2 contribute to the cross sections.
A large strength is noticed for the transition  
to the ground state of $^{18}$F (1$^{+}$, T=0). 
The strength of the Fermi transition, $B(F)$, is defined as
\begin{equation}
B(F) = \frac{1}{2J_i +1} |<f|| \sum_{k} t_{-}^{k} ||i>|^2.
\end{equation}
The value of the $B(F)$ for the transition to the isobaric analog state (IAS), $^{18}$F (0$^{+}$, 1.04 MeV, T=1) is equal to 2.
Note that the isospin of the final states are T =1, 2 and only T=2 for the transitions to $^{18}$O and $^{18}$N, respectively. 
The transition strengths for $^{18}$O and $^{18}$N are, therefore, suppressed compared to the strength for $^{18}$F.

\begin{figure}[htbp]
\begin{minipage}{0.45\hsize}
\begin{center}
\hspace*{1cm}
\includegraphics[scale=0.55]{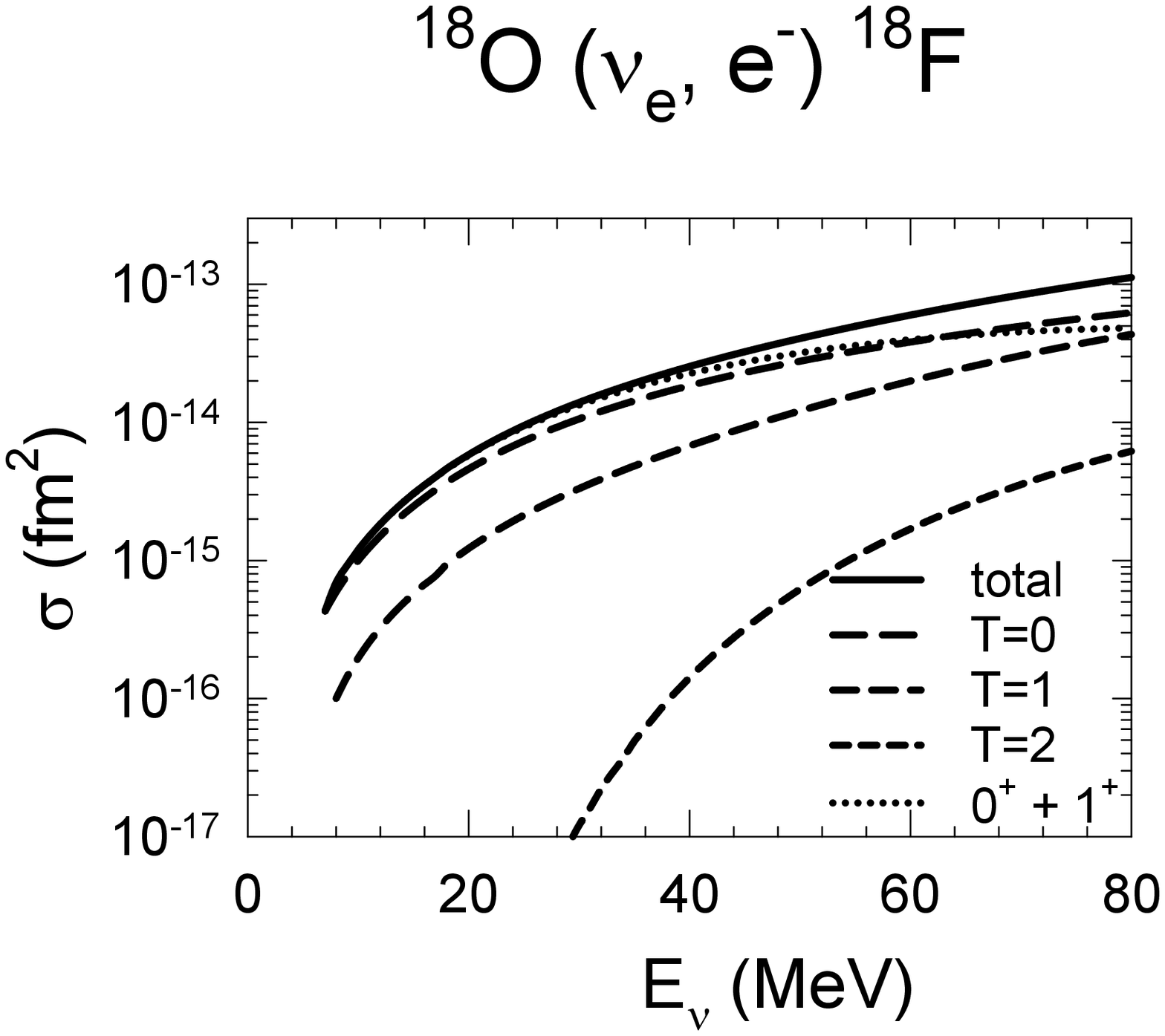}
\end{center}
\end{minipage}
\vspace*{-3.5cm}
\caption{
\small 
Calculated cross sections for $^{18}$O ($\nu_e$, e$^{-}$) $^{18}$F. Solid curve shows the total cross section. 
Long-dashed, dashed and short-dashed curves denote its components of transitions to the states with T=0, 1 and 2, respectively.
Dotted curve shows the sum of cross sections for the GT (1$^{+}$) and Fermi (0$^{+}$) transitions.} 
\label{crosf18}
\end{figure}

\subsection{Reaction cross sections for $^{18}$O}
In this subsection, reaction cross sections for $^{18}$O ($\nu_e$, e$^{-}$) $^{18}$F, $^{18}$O ($\bar{\nu}_e$, e$^{+}$) $^{18}$N and $^{18}$O ($\nu$, $\nu'$) $^{18}$O are evaluated by shell-model calculations in $p$-$sd$ shell with the use of SFO-tls.
Configurations up to 2p-2h (3p-3h) excitations are included for positive (negative) parity states. 
The cross sections are obtained by using the multipole expansion of the weak hadronic currents,
\begin{equation}
J_{\mu}^{C_{\mp}} = J_{\mu}^{V_{\mp}} + J_{\mu}^{A_{\mp}} 
\end{equation}
for charged-current reactions ($\nu_e$, e$^{-}$) and ($\bar{\nu}_e$, e$^{+}$), and 
\begin{equation}
J_{\mu}^{N} = J_{\mu}^{A_3} + J_{\mu}^{V_3} -2 \sin^2\theta_{W} J_{\mu}^{\gamma}
\end{equation}
for neutral-current reactions, ($\nu$, $\nu'$) and ($\bar{\nu}$, $\bar{\nu}'$), where $J_{\mu}^{V}$ and $J_{\mu}^{A}$ are vector and axial-vector currents, respectively, and $J_{\mu}^{\gamma}$ is the electromagnetic vector current with $\theta_{W}$ the Weinberg angle. 
The reaction cross sections are given as the sum of the matrix elements of the Coulomb, longitudinal, and transverse electric and magnetic multipole operators for the vector and axial-vector currents \cite{Wal,SC2006}. 
Here, all the transition matrix elements with the multipolarities up to $\lambda$ =4 are taken into account with the use of harmonic oscillator wave functions.
The quenching factor for $g_A$ determined in Sect. 2.1, $q$=0.88, is used for all the multipoles.
Calculated results are shown in Fig. 2 as functions of neutrino energy $E_{\nu}$.
We note that the contributions from the GT (1$^{+}$) and Fermi (0$^{+}$) transitions are dominant at $E_{\nu} <$ 40 MeV, and 
the transitions to the states with isospin T=0 are most important. 
The T=0 component is mostly the GT transitions, and
the T=1 component is dominantly the Fermi transition to the IAS state.
Spin-dipole transitions are the dominant contributions to the T=2 component.
The calculated total cross section is found to be larger than that for $^{16}$O as will be shown in Sect. 2.3.

\begin{figure}[htbp]
\vspace*{-0.5cm}
\hspace{-8mm}
\begin{minipage}{0.45\hsize}
\begin{center}
\includegraphics[scale=0.37]{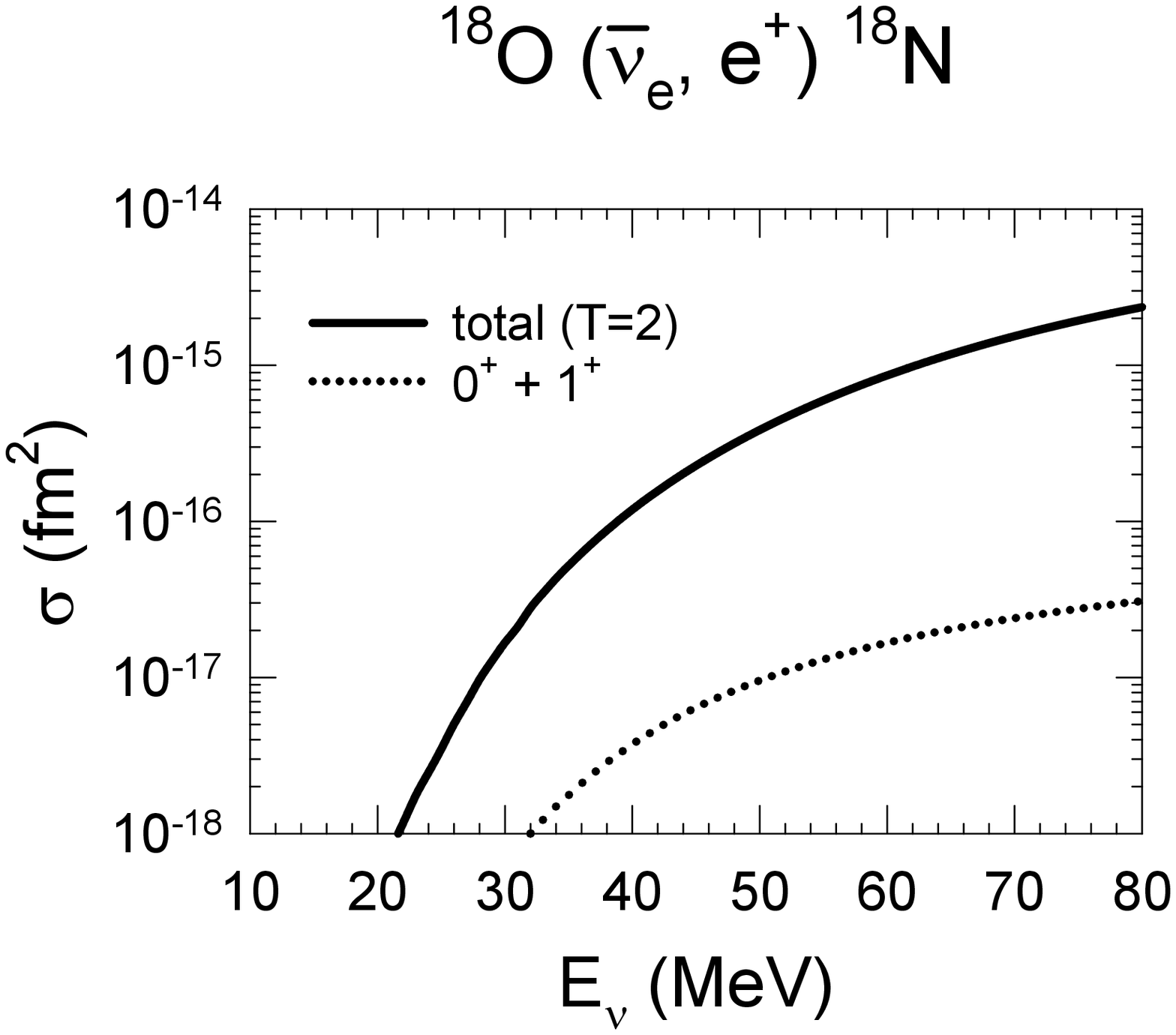}
\end{center}
\end{minipage}
\hspace{5mm}
\begin{minipage}{0.45\hsize}
\begin{center}
\includegraphics[scale=0.37]{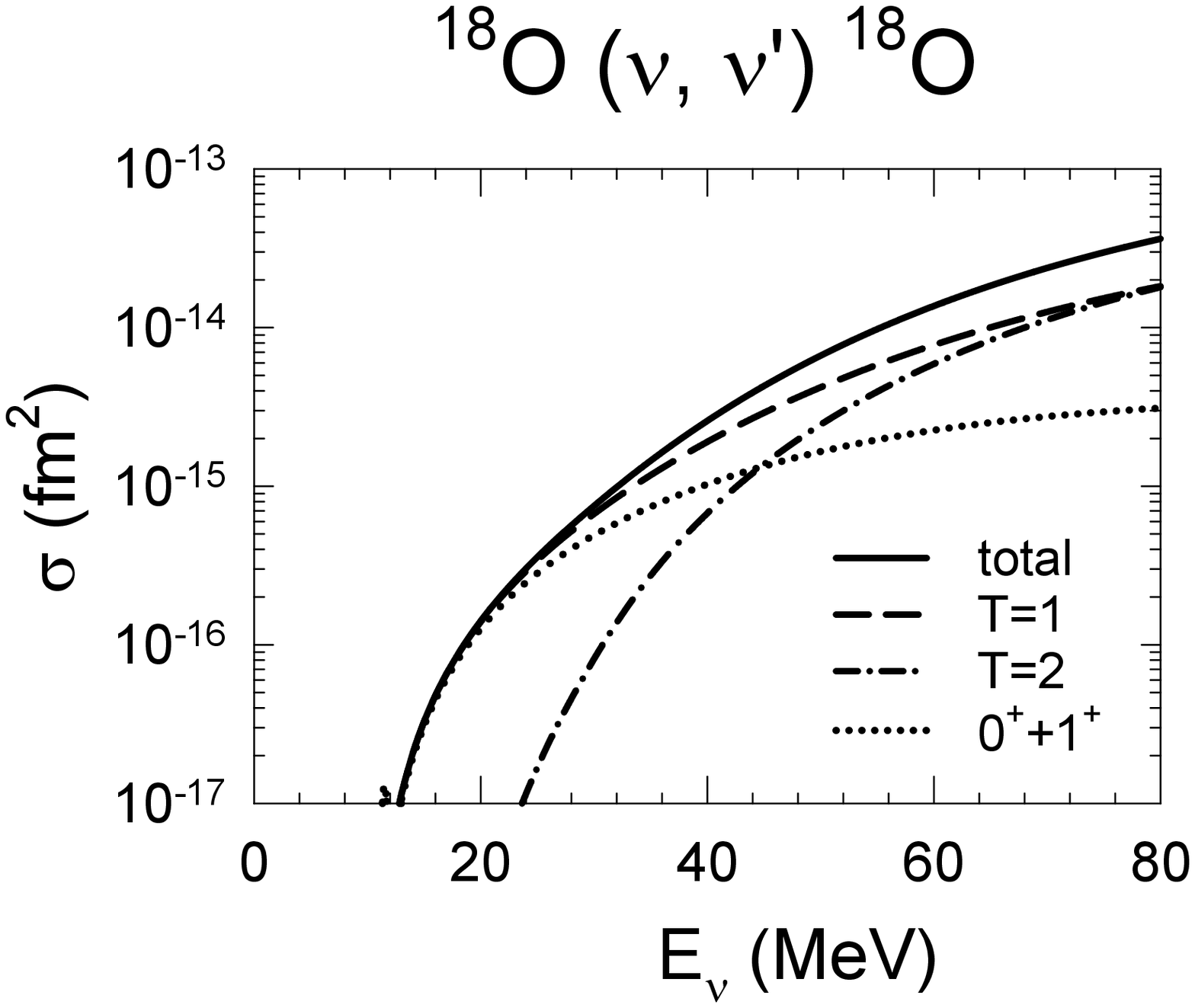}
\end{center}
\end{minipage}
\vspace{-2cm}
\caption{\small 
Calculated reaction cross sections for $^{18}$O ($\bar{\nu}_e$, e$^{+}$) $^{18}$N (left panel) and $^{18}$O ($\nu$, $\nu'$) $^{18}$O (right panel).
Total cross section and the sum of cross sections for 0$^{+}$ and 1$^{+}$ multipoles are shown.  
Final states of $^{18}$N have isospin T=2, while those of $^{18}$O have T=1 and 2.}
\label{croson18}
\end{figure}

Cross sections for ($\bar{\nu}_e$, e$^{+}$) and ($\nu$, $\nu'$) reactions are also evaluated by the multipole expansion method, and the results are shown in Fig. 3.
The calculated cross sections are smaller than those for the ($\nu_e$, e$^{-}$) reaction by more than one-order of magnitude 
because there are no 
first-order allowed GT transitions to T=0 states, that are present in the ($\nu_e$, e$^{-}$) reaction channel.
Note also that the transition to the IAS state (0$^{+}$, T=1) does not exist in the ($\bar{\nu}_e$, e$^{+}$) and ($\nu$, $\nu'$) reactions.
A large threshold energy for the ($\bar{\nu}_e$, e$^{+}$) reaction further suppresses the magnitude of the cross section.

\subsection{Comparison with reaction cross sections for $^{16}$O}
Now, we compare the cross sections for $^{18}$O with those for $^{16}$O.
The calculated total cross sections for $^{18}$O ($\nu_e$, e$^{-}$) $^{18}$F are compared with those for $^{16}$O ($\nu_e$, e$^{-}$) $^{16}$F in Fig. 4 (left panel).
Here, the quenching factor for $g_A$ for $^{16}$O is taken to be $q$ =0.68 (dotted curve) for the transitions to the first 0$^{-}$, 1$^{-}$ and 2$^{-}$ states in $^{16}$F \cite{PTEP2022}, which was determined by fitting to the experimental $\mu$-capture rates (see Ref. \cite{PTEP2022} for the details).
For transitions to other states in $^{16}$F, the same value as in Ref. \cite{SC2018} ($q$=0.95), which is consistent with the total experimental $\mu$-capture rate to unbound states, is used (see Table B1 of Ref. \cite{PTEP2022}).
The cross section for $^{16}$O is reduced by about 50$\%$ at $E_{\nu}<$ 20 MeV and 30$\%$ (20$\%$) around $E_{\nu}$ = 30 (40) MeV compared with that in Ref. \cite{SC2018}, where $q$=0.95 is adopted for all the transitions.
The cross section for $^{18}$O is larger than that for $^{16}$O due to the large contribution from the GT transitions in $^{18}$O.
Even if 0.205$\%$ for the isotope abundance of $^{18}$O is taken into account, the cross section for $^{18}$O is still larger at low neutrino energies, $E_{\nu}\leq$ 25 MeV.
Calculated cross sections for $^{18}$O with 0.205$\%$ abundance are consistent with those obtained in Ref. \cite{Haxton} (see Fig. 1).
They are also comparable to the $\nu_{e}$-$e^{-}$ elastic cross sections at $E_{\nu}\geq$ 20 MeV, and even larger at $E_{\nu}>$ 50 MeV.
Calculated cross sections for $^{16}$O ($\bar{\nu_e}$, e$^{+}$) $^{16}$N are also shown in Fig. 4 (right panel). The cross section obtained with $q$=0.68 (solid curve) for the transition to the first 0$^{-}$, 1$^{-}$ and 2$^{-}$ states in $^{16}$N is reduced by about half at $E_{\nu}\leq$ 20 MeV compared with that obtained with $q$=0.95 \cite{SC2018}. 
We also notice by comparing to Fig. 3 that the cross section for $^{16}$O is larger than that for $^{18}$O by one-order of magnitude in the ($\bar{\nu_e}$, e$^{+}$) channel.

\begin{figure}[htbp]
\hspace{-10mm}
\begin{minipage}{0.45\hsize}
\begin{center}
\includegraphics[scale=0.40]{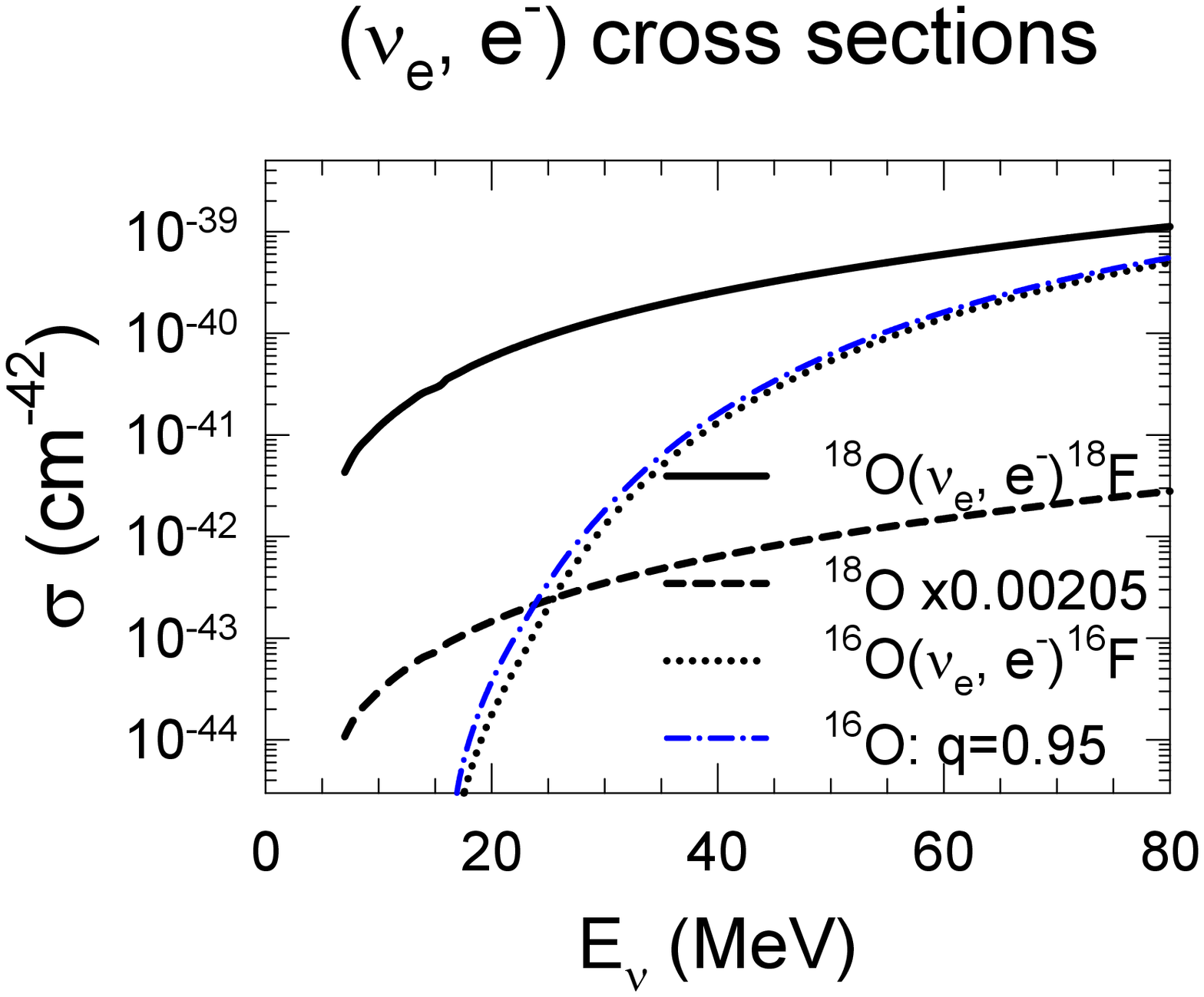}
\end{center}
\end{minipage}
\hspace{8mm}
\begin{minipage}{0.45\hsize}
\begin{center}
\includegraphics[scale=0.40]{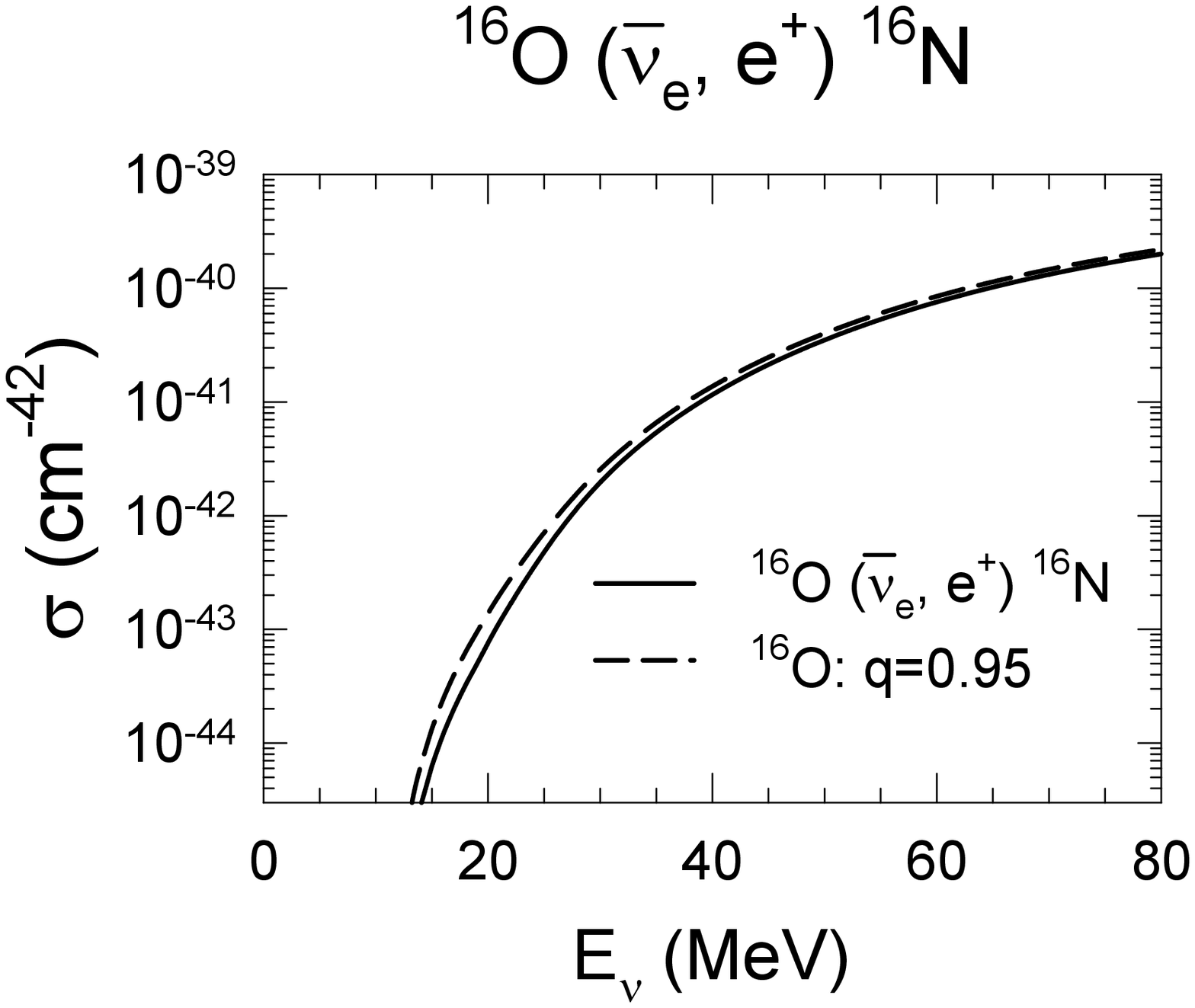}
\end{center}
\end{minipage}
\vspace{-2.5cm}
\caption{\small 
(Left) Comparison of ($\nu_e$, e$^{-}$) total cross sections for $^{18}$O and $^{16}$O. 
Dotted curve is obtained by shell model with the use of $q$=0.68 for the quenching of $g_A$ for the transitions to the first 0$^{-}$, 1$^{-}$ and 2$^{-}$ states of $^{16}$F.
Dash-dotted curve is taken from Ref. \cite{SC2018}, where $q$=0.95 is used.
Dashed curve denotes the cross section for $^{18}$O, where 0.205$\%$ isotope abundance of $^{18}$O is taken into account.
(Right) Cross sections for $^{16}$O ($\bar{\nu_e}$, e$^{+}$) $^{16}$N obtained by shell-model calculations. 
Solid curve is obtained with $q$=0.68 for the transitions to the first 0$^{-}$, 1$^{-}$ and 2$^{-}$ states of $^{16}$N, while dashed curve is obtained with $q$=0.95 \cite{SC2018}.
}
\label{1816O}
\end{figure}

\begin{figure}[htbp]
\vspace*{-2cm}
\hspace*{-2.0cm}
\begin{minipage}{0.45\hsize}
\begin{center}
\includegraphics[scale=0.40]{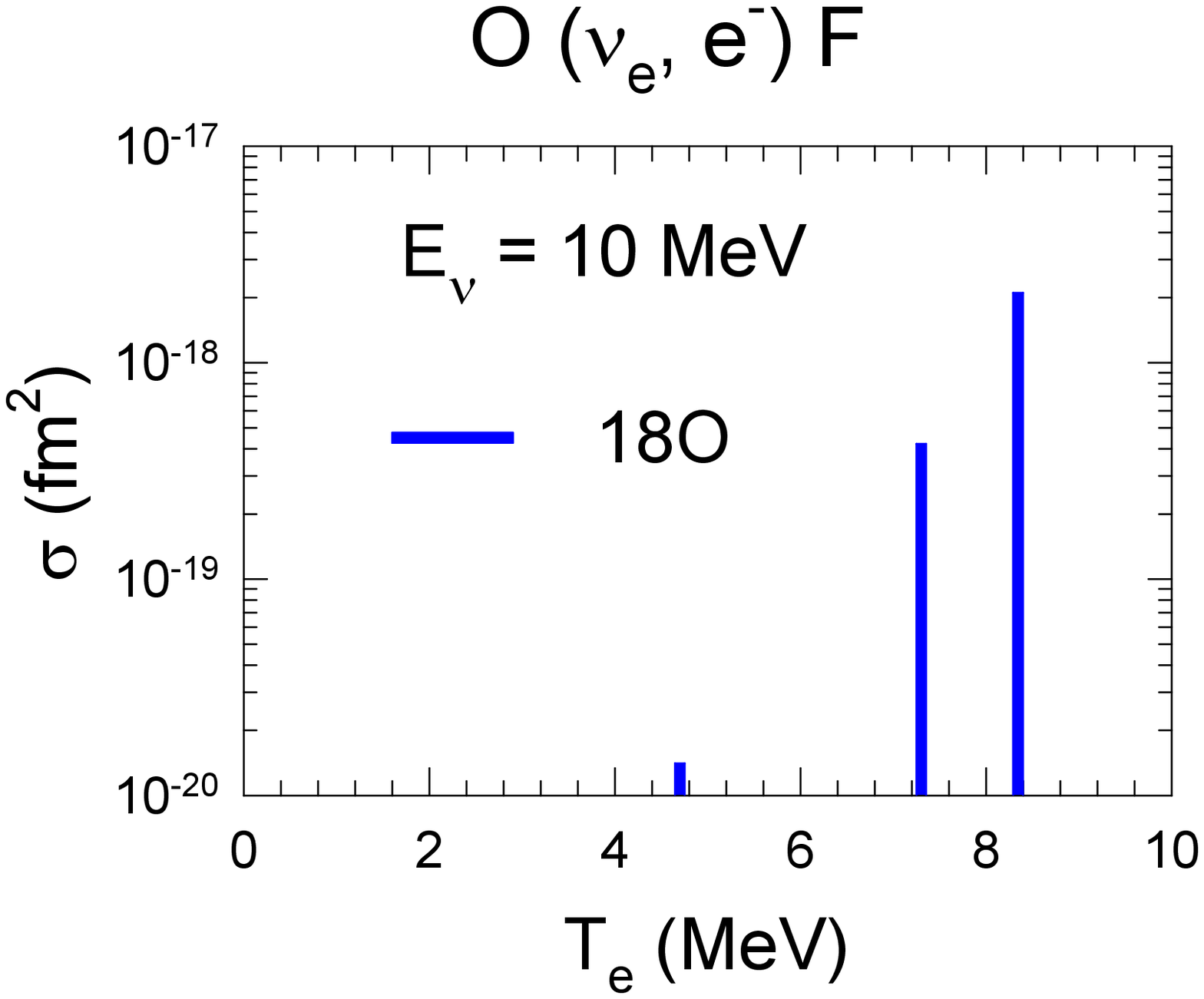}
\end{center}
\end{minipage}
\hspace{-15mm}
\begin{minipage}{0.45\hsize}
\begin{center}
\hspace*{-0.7cm}
\includegraphics[scale=0.40]{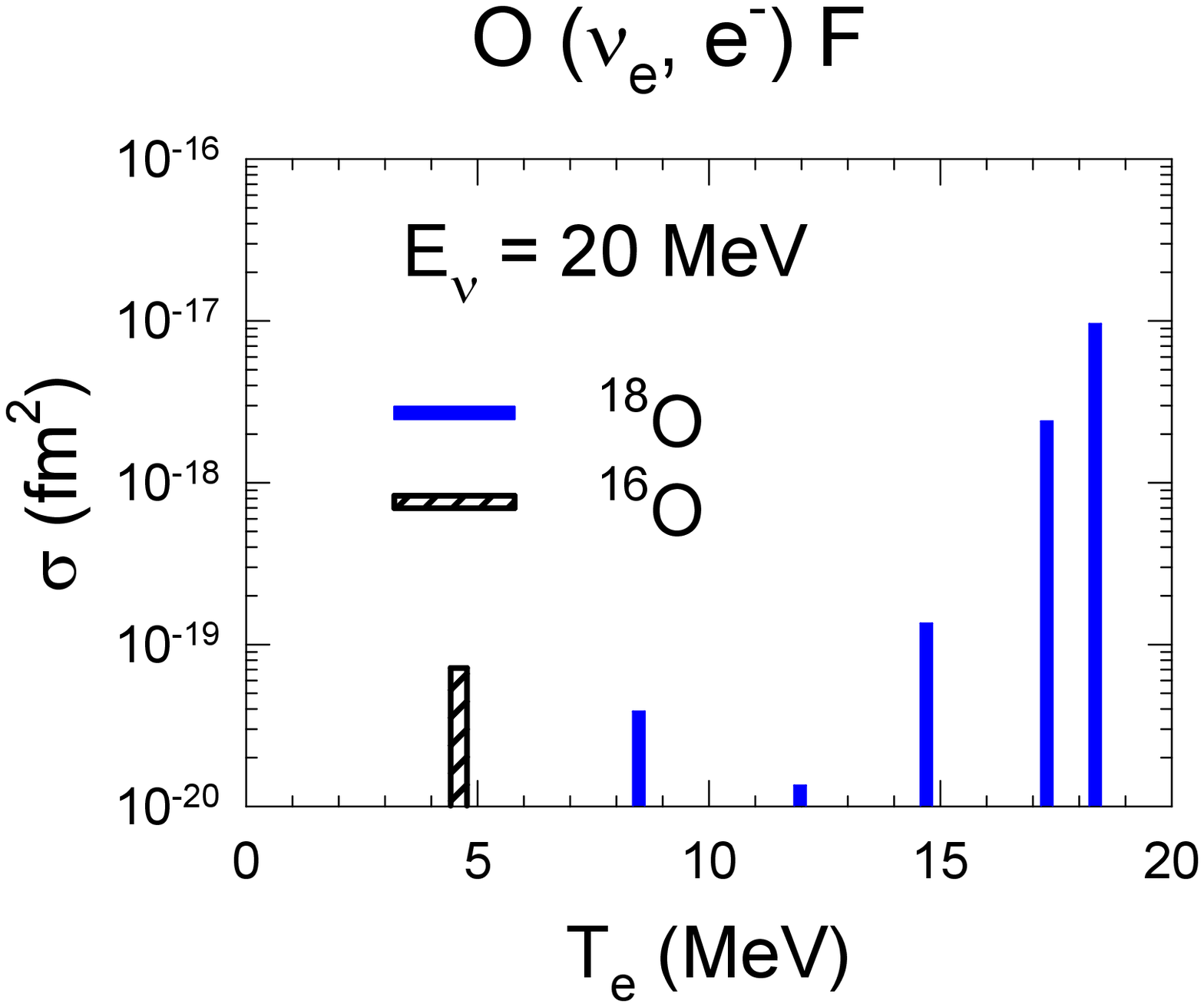}
\end{center}
\end{minipage}
\hspace{-15mm}
\vspace*{-1.0cm}
\begin{minipage}{0.45\hsize}
\begin{center}
\vspace*{-1.0cm}
\hspace*{-0.7cm}
\includegraphics[scale=0.40]{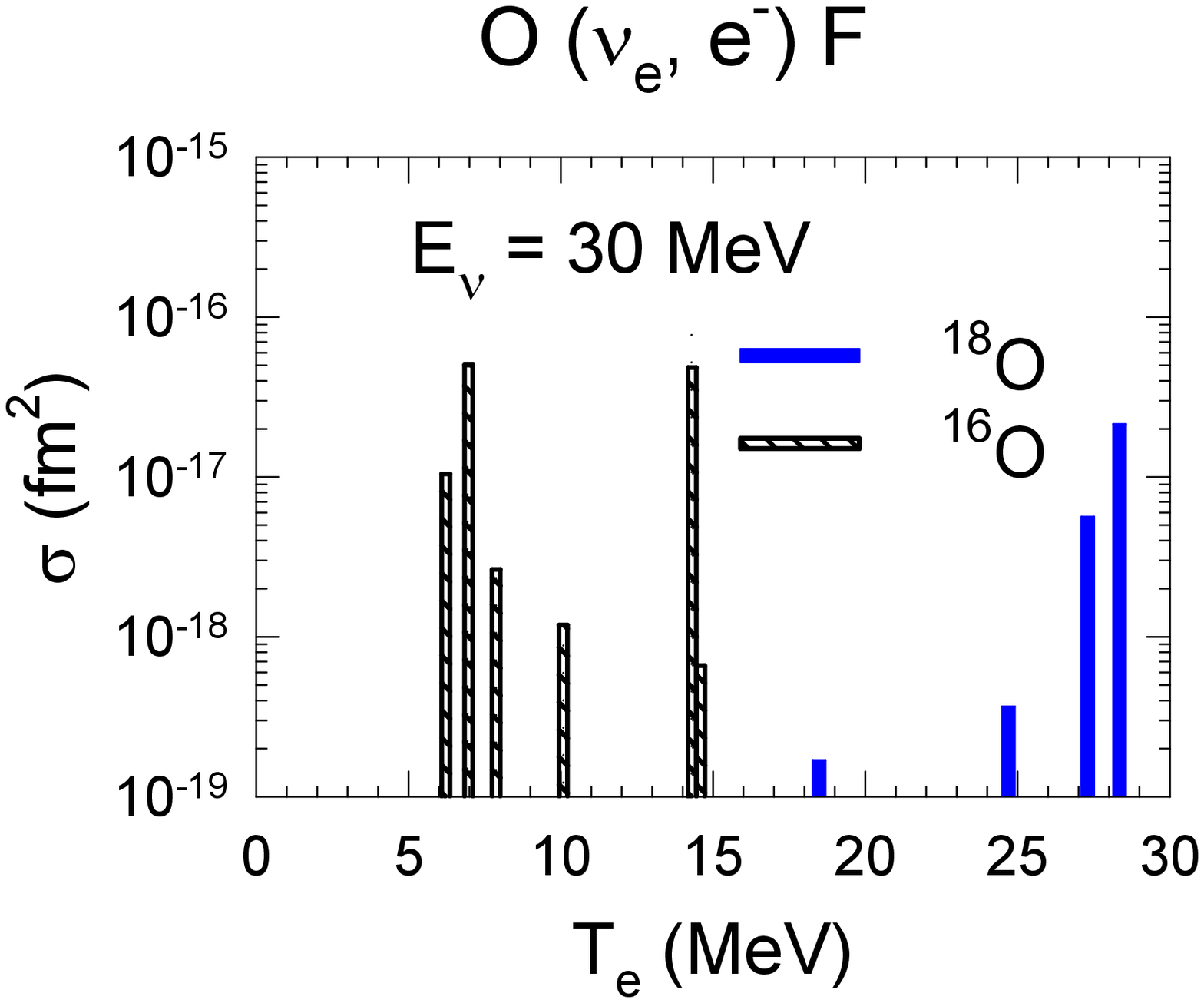}
\end{center}
\end{minipage}
\begin{minipage}{0.45\hsize}
\begin{center}
\vspace*{-1.0cm}
\hspace*{-1.2cm}
\includegraphics[scale=0.40]{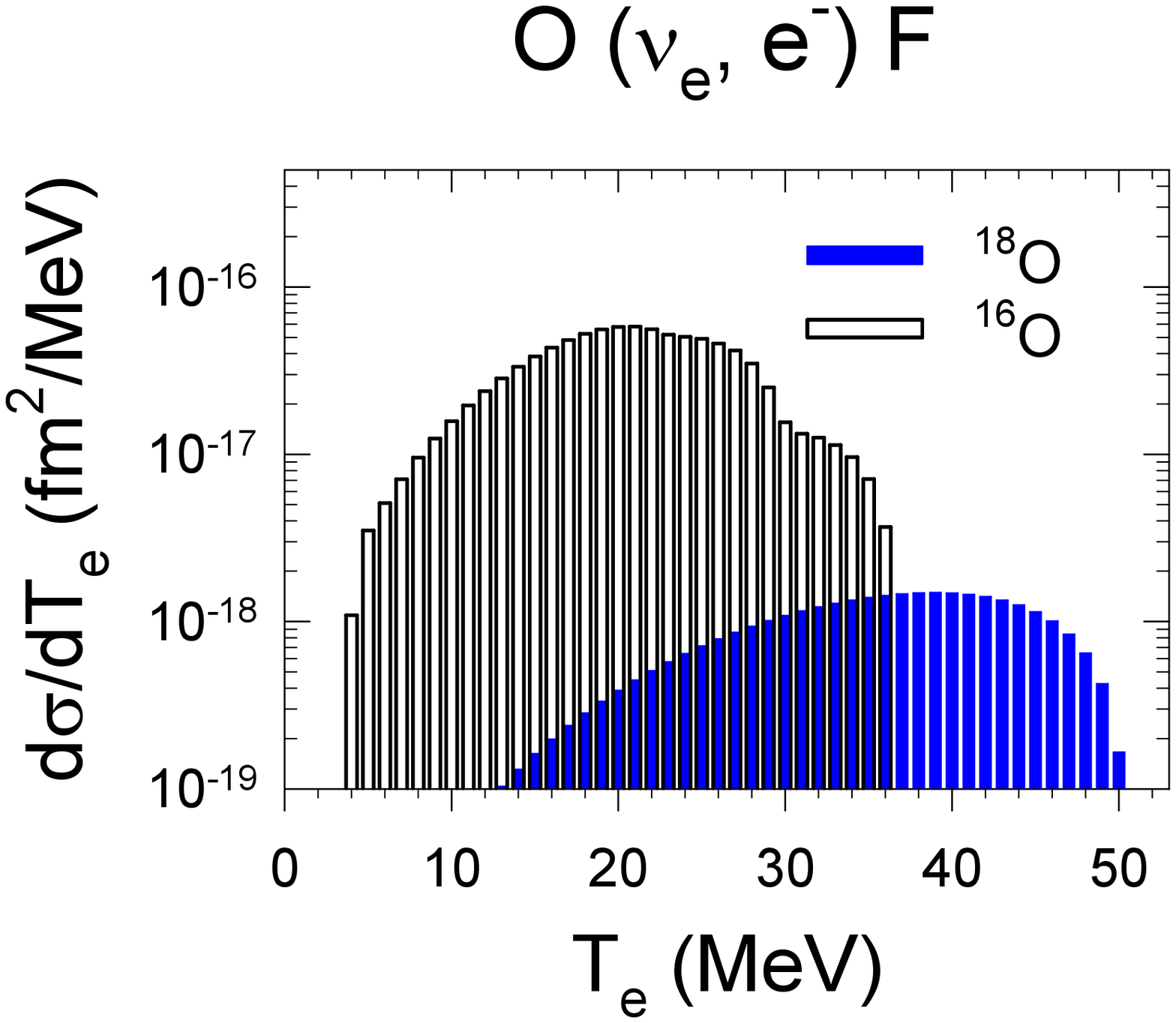}
\end{center}
\end{minipage}
\vspace{-1cm}
\caption{\small 
Contributions from $^{18}$O and $^{16}$O in water to O ($\nu_e$, e$^{-}$) F cross sections as functions of emitted electron kinetic energy, $T_e$.
Cases for $E_{\nu}$ = 10, 20 and 30 MeV as well as the one folded over the DAR $\nu_e$ spectrum are shown.
}
\label{1816OEn}
\end{figure}

Contributions to the ($\nu_e$, e$^{-}$) cross sections from $^{18}$O and $^{16}$O in water are compared for $E_{\nu}$ = 10, 20 and 30 MeV in Fig. 5 as functions of kinetic energy of emitted electron, $T_e$.
Threshold energy of the detector is taken to be 5 MeV.
As the threshold energy of for $^{16}$O ($\nu_e$, e$^{-}$) $^{16}$F reaction is 15.42 MeV, there are no contributions from $^{16}$O below $E_{\nu}$ = 20 MeV.
There are contributions from $^{18}$O only at $E_{\nu}$ = 10 MeV.
Note that the threshold energy for $^{18}$O ($\nu_e$, e$^{-}$) $^{18}$F reaction is as low as 1.66 MeV. 
Contributions at $T_e$ = 8.34 and 7.30 MeV come from the GT transition to the ground state of $^{18}$F and the IAS, respectively. 
For $E_{\nu}$ = 20 MeV, a large contribution from $^{18}$O caused by the transitions to the GT and IAS states is seen at $T_e$ =18.34 and 17.30 MeV, while the transition to the ground state of $^{16}$F (0$^{-}$) at $T_e$ =4.59 MeV is the only contribution from $^{16}$O.  
For $E_{\nu}$ = 30 MeV, contributions from $^{16}$O become large at $T_e$ = 7-10 and $\approx$14.3 MeV, but those from $^{18}$O are also noticed in a different energy region at $T_e$ =27-28 MeV though their magnitude is smaller compared to $^{16}$O.
Contributions from $^{18}$O are thus expected to be found at higher $T_e$ region.
In particular, they can be observed exclusively below $E_{\nu}\approx$ 20 MeV. 
The cross sections folded over the decay-at-rest (DAR) $\nu_e$ spectrum are also shown in Fig. 5.
There are no contributions from $^{16}$O at $T_e >$37 MeV, while those from $^{18}$O are found at $T_e$ = 37-50 MeV. 
Thus the contributions from $^{18}$O can be  exclusively observed at the higher electron energy region.  
Note also that $\nu_e$-induced reactions on $^{18}$O have almost isotropic angular distributions, while those on $^{16}$O are backward-peaked \cite{Haxton}.  
An experiment with DAR $\nu_e$ should be able to measure the cross sections of $^{18}$O and $^{16}$O separately, and test the present model calculations \cite{COHERENT}.  

\begin{figure}[htbp]
\vspace*{-0.5cm}
\hspace{-7mm}
\begin{minipage}{0.45\hsize}
\begin{center}
\includegraphics[scale=0.38]{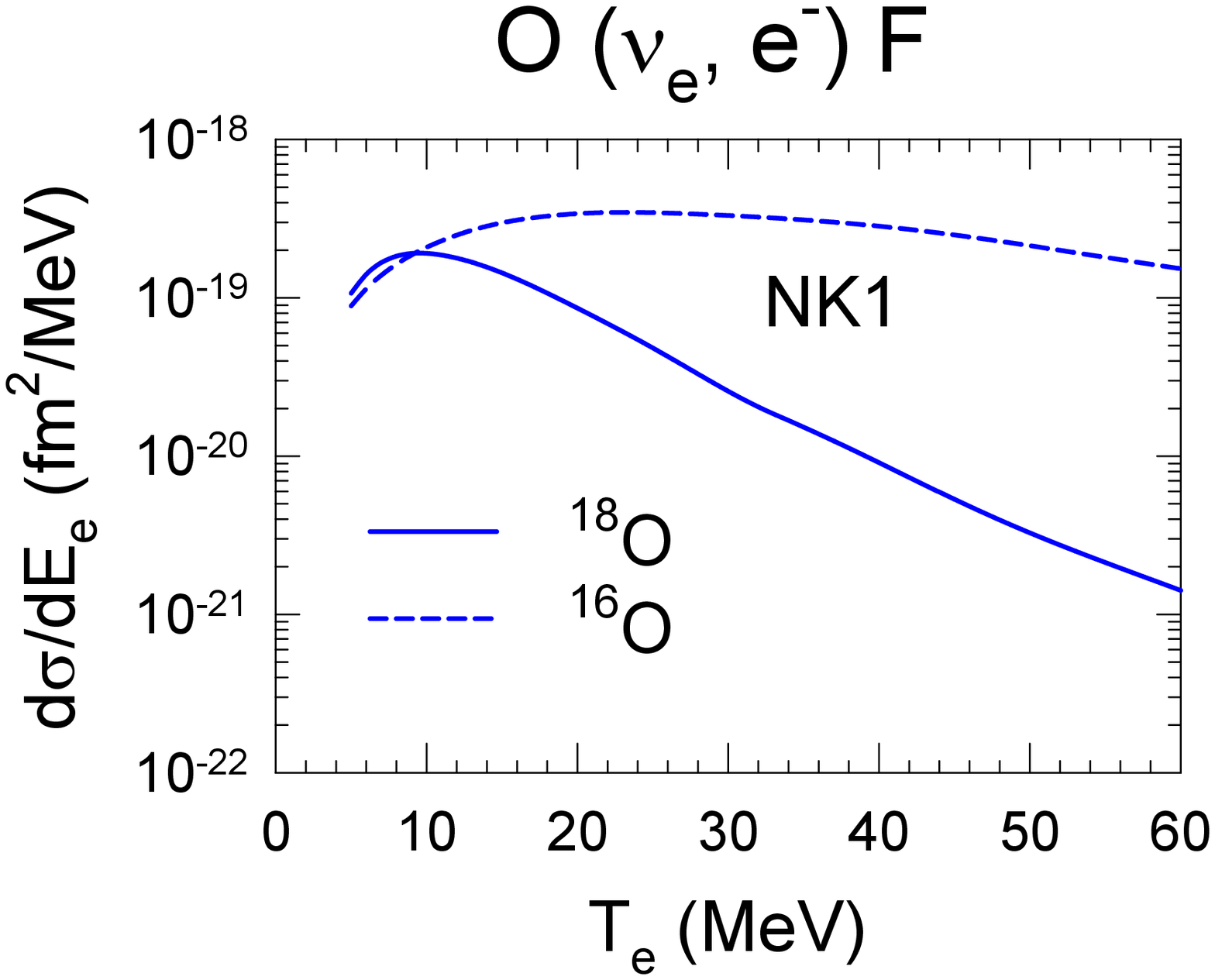}
\end{center}
\end{minipage}
\hspace*{10mm}
\begin{minipage}{0.45\hsize}
\begin{center}
\includegraphics[scale=0.38]{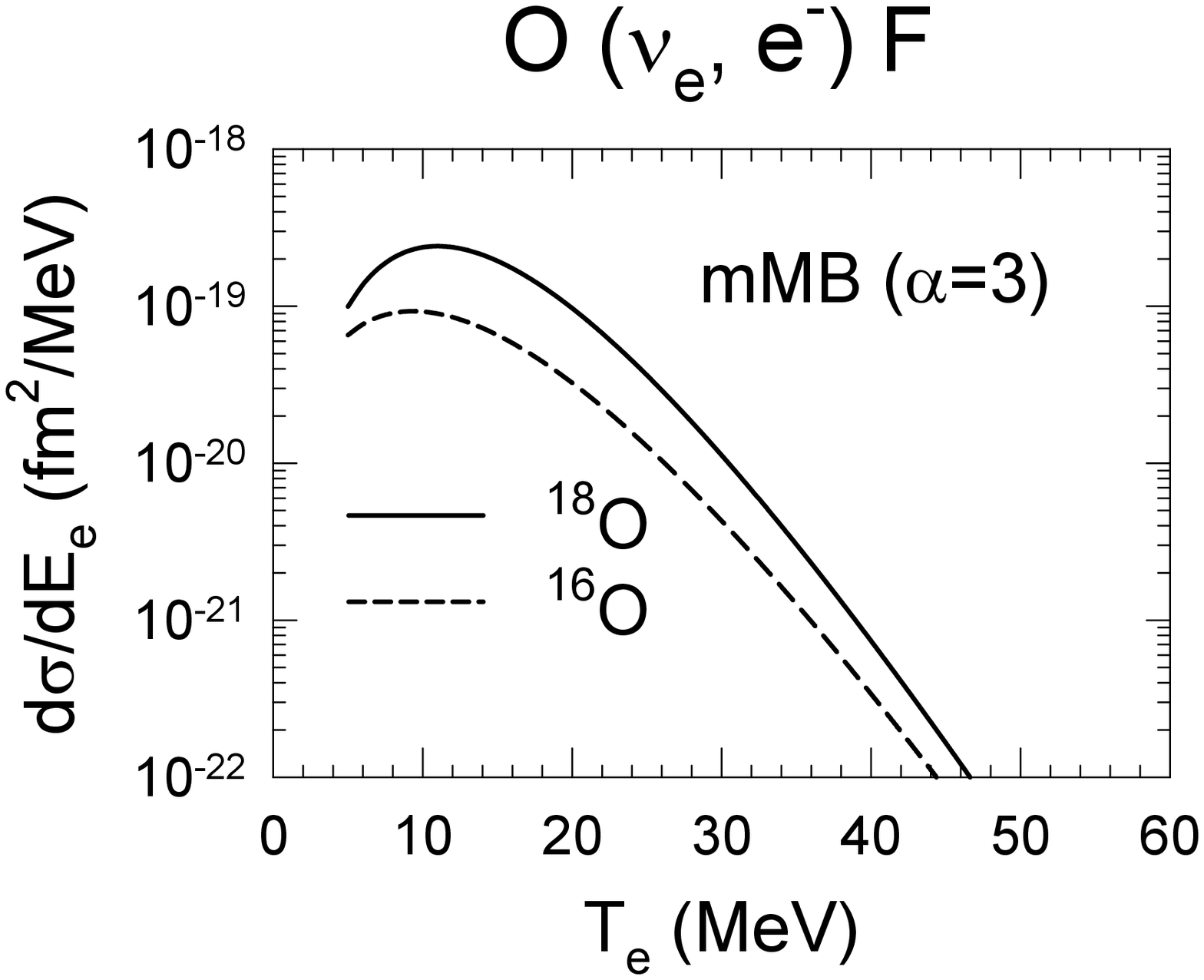}
\end{center}
\end{minipage}
\vspace{-2cm}
\caption{\small 
Cross sections for ($\nu_e$, e$^{-}$) reactions on $^{18}$O and $^{16}$O as functions of emitted electron energies, $T_e$. 
Neutrino spectra of a normal supernova, NK1 \cite{Nakazato2018,Naka2013}, and modified Maxewll-Boltzmann (mMB) distributions with $\alpha$=3 and neutrino average energy of 10 MeV are adopted for left and right panels, respectively.
} 
\label{ratio}
\end{figure}

\section{Effects of $^{18}$O mixture in water on supernova $\nu$ detection}
\subsection{Cross sections folded over supernova neutrino spectra}

In order to make an estimate for the supernova neutrino event rate, the folding effects of the ($\nu_e$, e$^{-}$) cross sections over supernova neutrino spectra are investigated. 
Cross sections for ($\nu_e$, e$^{-}$) reactions on $^{18}$O and $^{16}$O folded over supernova neutrino spectra as functions of emitted electron energies $T_e$ are shown in Fig. 6. 
The case of neutrino spectra of a normal supernova obtained by simulations of supernova explosions, denoted as NK1 \cite{PTEP2022,Nakazato2018}, and 
the case for a modified Maxwell-Boltzmann (mMB) distribution with neutrino average energy of 10 MeV are shown.
The NK1 model is one of the spectral models provided in Supernova Neutrino Database \cite{Naka2013} and its progenitor has the mass of $M=20M_\odot$ and the metallicity of $Z=0.02$.
For the mMB, the following parametrization is adopted with $\alpha$ =3 \cite{KRJ}:
\begin{equation}
f(E_{\nu}) = \frac{(\alpha +1)^{\alpha +1}}{\Gamma(\alpha +1) <E_{\nu}>^{\alpha +1}} E_{\nu}^{\alpha} \exp( -\frac{(\alpha +1)E_{\nu}}{<E_{\nu}>})
\end{equation}
where $<E_{\nu}>$ is average neutrino energy.
As $T_e$ increases, the cross sections folded over the NK1 spectra decrease more slowly compared with those of the mMB spectra for both $^{18}$O and $^{16}$O.
This is due to a large high energy component that remains up to $E_{\nu}\approx$ 100 MeV in the spectra of NK1.
Higher energy electrons are expected to be emitted more for the NK1 spectra.  
The cross sections for $^{18}$O and $^{16}$O are comparable up to $T_e\approx$10 MeV, but those for $^{16}$O are more enhanced compared to $^{18}$O at higher $T_e$ regions for the NK1 case.
In case of the mMB spectra, on the other hand, the cross sections for $^{18}$O and $^{16}$O are comparable up to $T_e\approx$ 50 MeV, but their magnitudes decrease more rapidly as $T_e$ increases.   
This indicates that contributions from $^{18}$O are non-negligible 
and can affect the count rate for supernova $\nu$ events in water Cherenkov detectors.
As the high energy component contributes more to the events on $^{16}$O than on $^{18}$O for the NK1 spectra, the $^{18}$O admixture can affect the event rate less than the case for the mMB spectra.

\subsection{Supernova $\nu$ event rates in water Cherenkov detectors}

In this subsection, we estimate the event rates for supernova $\nu$ detection by using the neutrino spectra given by NK1, mMB and Fermi distributions. 
For this purpose, a supernova at 10 kpc and the detection at 32 kton water Cherenkov detector, Superkamiokande, are assumed.
Furthermore, cases with and without the Mikheyev-Smirnov-Wolfenstein (MSW) neutrino oscillations \cite{MSW} are considered.
The neutrino number spectrum for $\nu_e$ 
is given by \cite{Nakazato2018} 
\begin{equation}
\frac{dN_{\nu_e}(E_{\nu})}{dE_{\nu}} =  P \frac{dN_{\nu_e}^{0}(E_{\nu})}{dE_{\nu}} + (1-P) \frac{dN_{\nu_x}^{0}(E_{\nu})}{dE_{\nu}}
\end{equation}
where $P$ is the survival probability of $\nu_{e}$ and $\frac{d N_{\nu_{i}}^{0}}{d E_{\nu}}$ = $\frac{E_{\nu_{i,{\rm total}}}}{<E_{\nu_{i}}>}$ $f(E_{\nu_{i}})$ with $i=e$ and $x$ ($x$ =$\mu$, $\tau$, $\bar{\mu}$, and $\bar{\tau}$) are neutrino number spectra before the oscillations. $E_{\nu_{i,{\rm total}}}$ is the total energy emitted by $\nu_{i}$, $<E_{\nu_{i}}>$ is the average energy for $\nu_{i}$ and $f(E_{\nu_{i}})$ is the normalized neutrino spectrum for $\nu_{i}$. 
The value of $P$ is taken to be 0 and 0.32 for normal hierarchy (NH) and inverted hierarchy (IH), respectively \cite{Dighe}.
The similar estimations were done in Ref. \cite{Nakazato2018}, but only $^{16}$O was taken into account.

\begin{figure}[htbp]
\vspace*{-3cm}
\hspace{-15mm}
\begin{minipage}{0.45\hsize}
\begin{center}
\includegraphics[scale=0.38]{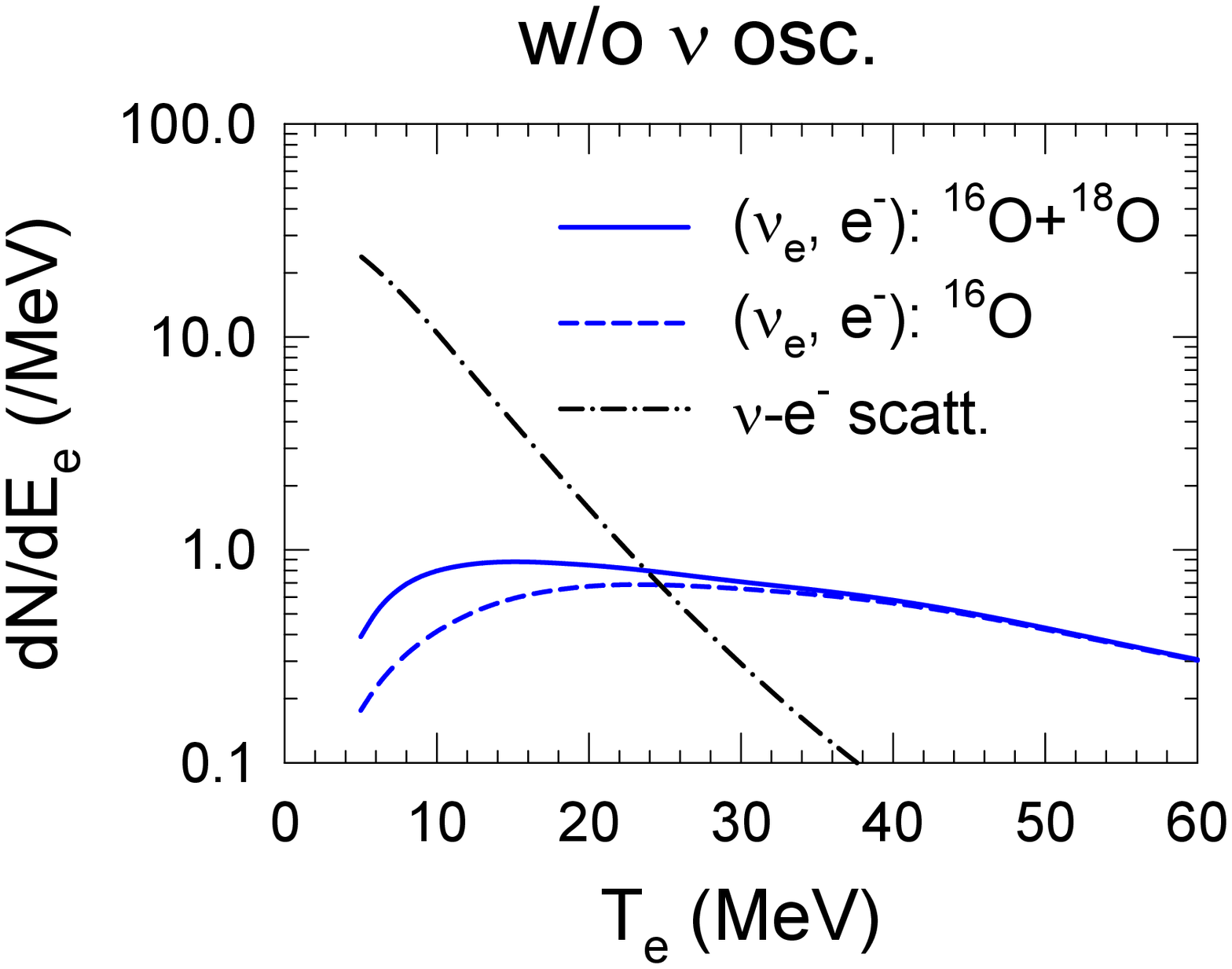}
\end{center}
\end{minipage}
\hspace{-15mm}
\begin{minipage}{0.45\hsize}
\begin{center}
\includegraphics[scale=0.38]{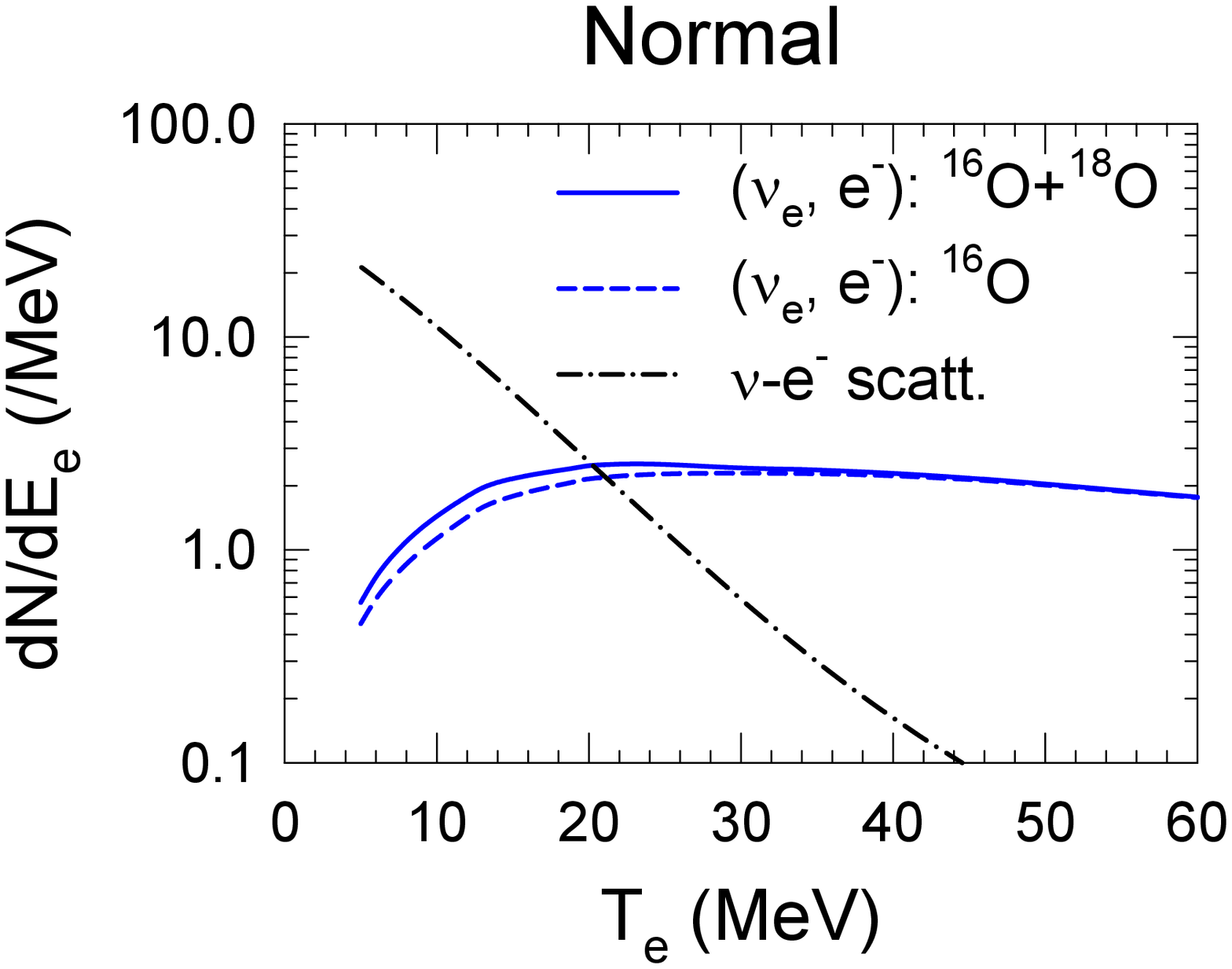}
\end{center}
\end{minipage}
\hspace{-15mm}
\vspace*{-1.0cm}
\begin{minipage}{0.45\hsize}
\begin{center}
\vspace*{-1cm}
\includegraphics[scale=0.38]{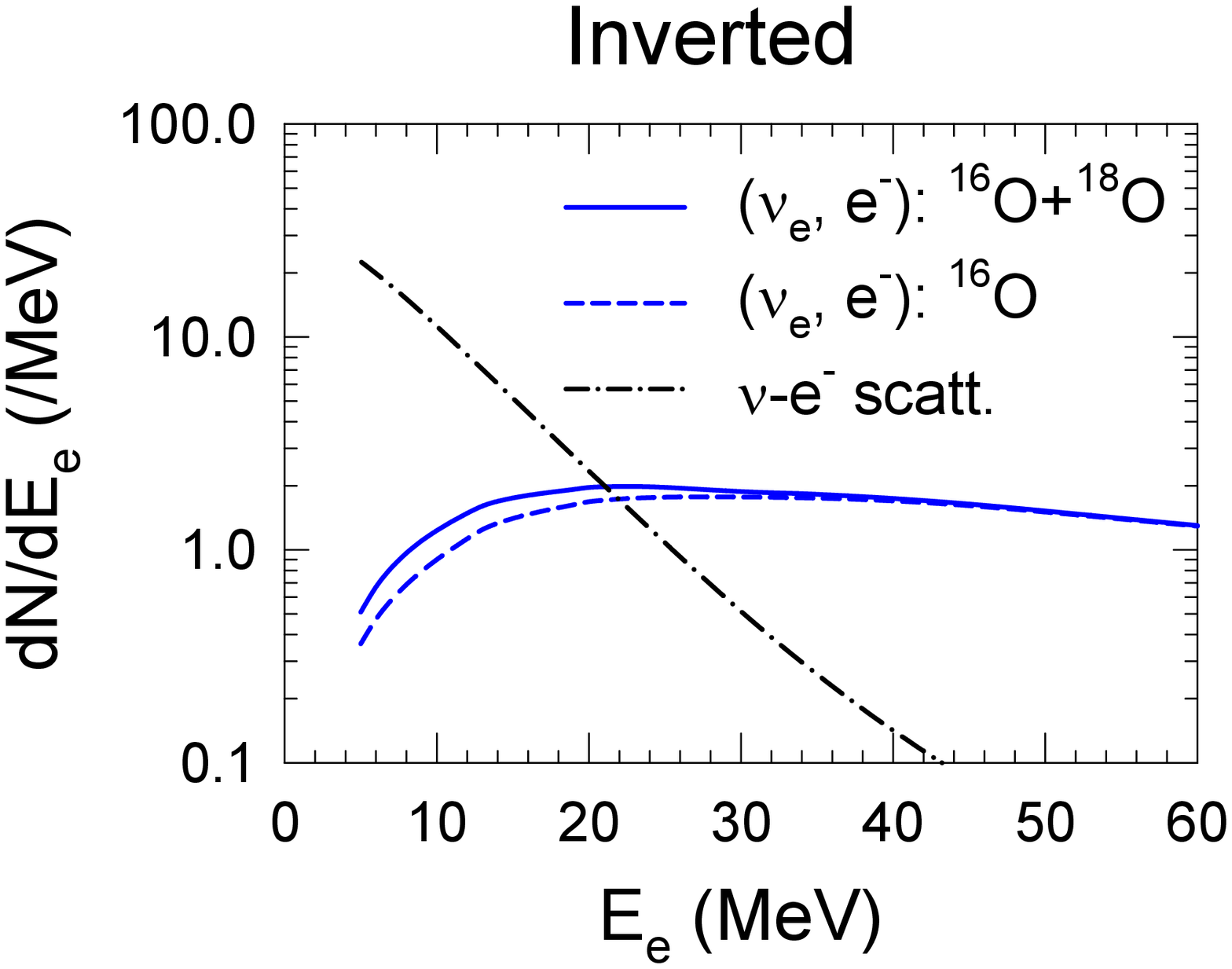}
\end{center}
\end{minipage}
\vspace{-1cm}
\caption{\small 
Event spectra for the supernova model, NK1 \cite{Nakazato2018,Naka2013}, as functions of emitted electron energy $T_e$ for the cases without $\nu$ oscillations (left), with the normal hierarchy hypothesis (center) and with the inverted hierarchy hypothesis (right). 
Solid and dashed curves correspond to ($\nu_e$, e$^{-}$) reactions on $^{16}$O + $^{18}$O with 0.205$\%$ and on $^{16}$O only, respectively. 
Dash-dotted curves correspond to elastic $\nu$+$e^{-}$ scattering.
}
\label{spectra}
\end{figure}

Here, we study possible effects of $^{18}$O mixture in water Cherenkov detectors on the event rates of supernova neutrinos.
Firstly, we consider the neutrino spectra of NK1 model, whose averaged neutrino energies for $\nu_e$, $\bar{\nu}_e$ and $\nu_x$ are 9.32, 11.1 and 11.9 MeV, respectively \cite{Naka2013}.
Event spectra for ($\nu_e$, e$^{-}$) reactions on $^{16}$O only and on $^{16}$O + $^{18}$O with 0.205$\%$ abundance as well as for elastic {\bf $\nu$-$e^{-}$} scattering are shown in Fig. 7 for the cases without $\nu$ oscillations and with oscillations for NH and IH.
For $\nu$-$e^{-}$ scattering, $\nu$ includes all flavors, $\nu_e$, $\bar{\nu}_e$, $\nu_{\mu}$, $\bar{\nu}_{\mu}$, $\nu_{\tau}$ and $\bar{\nu}_{\tau}$.
In case without the oscillation, the contribution from $^{18}$O enhances the spectra for $^{16}$O by about 60$\%$ at $T_{e}<$ 20 MeV.
A mild enhancement of the spectra by 20-30$\%$ from $^{18}$O admixture is noticed at $T_{e} \leq$25 MeV for the case with the oscillations.

\begin{table}[htbp]
\begin{center}
\begin{tabular}{|c|c|c|c|c|}
\hline
Neutrino spectra & Target& No oscillation & Normal & Inverted \\
\hline\hline
 NK1 \cite{Naka2013} & $\nu$+$e^{-}$ $\rightarrow$ $\nu$+$e^{-}$  & 140 & 157 & 156 \\
\hline
NK1 \cite{Naka2013} & $^{16}$O  & 36  & 156 & 118 \\
(9.32, 11.9) &$^{16}$O +$^{18}$O  & 42 & 165 & 126 \\
\hline
\hline
mMB \cite{Tambora} & $^{16}$O & 4 & 63 & 44 \\
(10.14, 12.89) & $^{16}$O +$^{18}$O & 11 & 76 & 55 \\
$\alpha$=(2.90, 2.39) & & & & \\
\hline
mMB & $^{16}$O & 14 & 50 & 39 \\
(12, 14.6) & $^{16}$O +$^{18}$O & 22 & 61 & 48 \\
$\alpha$=(2.9, 2.9) & & & & \\
\hline
mMB & $^{16}$O & 57 & 419 & 303 \\
(11.0, 15.8) & $^{16}$O +$^{18}$O & 70 & 439 & 321 \\
$\alpha$=(2, 2) & & & & \\
\hline
mMB & $^{16}$O & 6 & 67 & 48 \\
(11.0, 15.8) & $^{16}$O +$^{18}$O & 13 & 78 & 57 \\
$\alpha$=(3, 3) & & & & \\
\hline
mMB & $^{16}$O & 33 & 128 & 98 \\
(10.1, 12.6) & $^{16}$O +$^{18}$O & 45 & 144 & 112 \\
$\alpha$=(2, 2) & & & & \\
\hline
mMB & $^{16}$O & 3 & 16 & 12 \\
(10.1, 12.6) & $^{16}$O +$^{18}$O & 9 & 24 & 19 \\
$\alpha$=(3, 3) & & & & \\
\hline
\hline
\hline
Fermi \cite{Yoshida2005,Yoshida2008} & $^{16}$O & 13 & 110 & 78 \\
(3.5, 5) & $^{16}$O +$^{18}$O & 20 & 121 & 88 \\
\hline
Fermi & $^{16}$O & 7 & 30 & 23 \\
(3.2, 4) & $^{16}$O +$^{18}$O & 13 & 39 & 31 \\
\hline
\end{tabular}
\end{center}
\caption{
Results for expected event numbers 
for ($\nu_e$, e$^{-}$) reaction for a normal supernova \cite{Nakazato2018} (see text) are given in the row denoted as NK1. 
The threshold energy of the detector is taken to be 5 MeV.
Cases without neutrino oscillations, with the MSW oscillations with normal and inverted hierarchies are given for pure $^{16}$O and $^{16}$O with $^{18}$O mixture. Numbers in the bracket denote ($<E_{\nu_e}>$, $<E_{\nu_x}>$) in units of MeV.  
Expected event numbers for elastic $\nu$-$e^{-}$ scattering for the NK1 spectra are given in the first row \cite{Nakazato2018},  
where, $\nu$ includes all flavors.
Neutrino spectra of modified Maxwell-Boltzmann (denoted as mMB) and Fermi-Dirac distributions are also used for the estimation of the event numbers. 
Numbers in the brackets for the mMB distributions denote ($<E_{\nu_e}>$, $<E_{\nu_x}>$) in units of MeV and ($\alpha$ for $\nu_e$, $\alpha$ for $\nu_x$).  Those for the Fermi-Dirac distributions denote ($T_{\nu_e}$, $T_{\nu_x}$) in units of MeV.   }
\end{table}

Expected event numbers estimated for the ($\nu_e$, e$^{-}$) reaction for pure $^{16}$O and for $^{16}$O with 0.205$\%$ mixture of $^{18}$O are shown in Table I for the case with and without the MSW matter oscillation effects \cite{Dighe,Naka2013}.
Both NH and IH are considered for the MSW oscillations.
The event numbers for elastic $\nu$-$e^{-}$ scattering are also given in Table I for comparison. 
Note that the event numbers for pure $^{16}$O in Table I are different from those in Ref. \cite{Nakazato2018}.
This is because we adopt the quenching factor for $g_A$ of $q=0.68$ for the transitions from $^{16}$O to the first 0$^{-}$, 1$^{-}$ and 2$^{-}$ states in $^{16}$F \cite{PTEP2022}, while it was taken to be $q =0.95$ in Ref. \cite{Nakazato2018}.

In case of the NK1 neutrino spectra, the event numbers are found to be enhanced by 17$\%$ and 6-7$\%$  
with the $^{18}$O admixture
for the case without the oscillation and  with the oscillation, respectively.
Note that the $^{18}$O mixture becomes more important at lower $\nu_e$ energy region. 
In case with the oscillations, energy of $\nu_e$ is higher than the case without the oscillations as high energy $\nu_x$'s are converted to $\nu_e$, and the effects of $^{18}$O mixture become less than the case without the oscillation. 
The change of event numbers due to the $^{18}$O mixture is found to be rather modest, by 6-9 counts, 
because of the characteristic behavior of the neutrino spectra of NK1, whose strengths remain in the high energy region, $E_{\nu}>$ 50 MeV. 
The high energy component is produced in the accretion phase of supernova explosions \cite{Naka2013}.   
Due to the enhancement by effects of the neutrino oscillations, the event numbers become comparable to those of the elastic $\nu$-$e^{-}$ scattering, which is particularly important to detect the direction of the supernova within 3 to 5 degrees~\cite{superK}.
However, the $^{18}$O admixture would not affect the accuracy of the direction since the effect is not large enough to cover up the contributions from the $\nu$-$e^{-}$ elastic scattering below $T_e\approx$20 MeV (Fig. 7). 
Furthermore, the $\nu$-$e^{-}$ elastic scattering has sharply forward-peaked angular distributions, while $\nu_e$-induced reactions on $^{18}$O and $^{16}$O have almost isotropic and mildly backward-peaked angular distributions, respectively.
The measurement of $\nu$-$e^{-}$ elastic scattering remains an important method to determine the direction of the neutrino source.

we consider neutrino spectra with analytic forms.
In the following, total neutrino energy of $3.2\times10^{52}$ erg/flavor is assumed.
Results for the neutrino spectra of modified Maxwell-Boltzmann (mMB) and Fermi-Dirac distributions are also given in Table I.
For mMB, the spectra of Ref. \cite{Tambora} obtained at 1016 ms for the postbounce time for a spherically symmetric supernova model, and those that satisfy a relation $<E_{\nu_x}>$/$<E_{\nu_e}>$ $\approx$1.22 \cite{KRJ} are used. 
While change of the expected event numbers with the $^{18}$O admixture is modest, by 7-13 counts, similar to the case of NK1, 
the percentage of increase is larger than that of NK1 for each oscillation case.
In particular for the spectra of Ref. \cite{Tambora}, it is 175\% for the case without the oscillations.
For the late phase as 1016 ms after the bounce, the emission of high-energy $\nu_e$ is suppressed, which is reflected in a large $\alpha$ value of spectrum, $\alpha=2.90$.
Thus, the impact of the $^{18}$O admixture is notable due to the reactions with low-energy neutrinos for the case without the oscillations.
In case with the oscillations, the event number is larger because, similar to the case of NK1, $\nu_x$'s have a spectrum with high average energy and small $\alpha$ value and are converted to $\nu_e$.

So as to look into the dependence on parameters, cases for ($<E_{\nu_e}>$, $<E_{\nu_x}>$) = (11.0, 15.8) and (10.1, 12.6) MeV are also given, where $\nu_e$ and $\nu_x$ have the same value of $\alpha$ by $\alpha=2$ and 3.
The event numbers increase as the neutrino energies increase or the value of $\alpha$ decreases. 
Increase of the expected event number due to the $^{18}$O admixture is again modest, by 6-20 counts, while the total expected event number for ($\nu_e$, e$^{-}$) reaction, denoted as $^{16}$O +$^{18}$O in Table I, are sensitive to the parameters, 9-439 counts.
The impact of the $^{18}$O admixture is more notable not only for lower average energy but also for larger $\alpha$ because the spectrum is more pinched (high-energy tail suppressed) for larger $\alpha$.

For the Fermi-Dirac distribution, spectra with ($T_{\nu_e}$, $T_{\nu_x}$) = (3.5, 5) MeV is adopted. 
$T_{\nu_x}$ is obtained to be 4.8-6.6 MeV in Ref. \cite{Yoshida2005} and 5.4$\pm$1.1 MeV in Ref. \cite{Yoshida2008} to avoid an overproduction of $^{11}$B abundance during Galactic chemical evolution.  
The case for ($T_{\nu_e}$, $T_{\nu_x}$) = (3.2, 4) MeV 
is also given.
Averaged neutrino energy is related to the temperature by $<E_{\nu}>$ = 3.15 $T_{\nu}$ for the Fermi distribution.
Corresponding averaged neutrino energies of ($T_{\nu_e}$, $T_{\nu_x}$) = (3.5, 5) and (3.2, 4) MeV are ($<E_{\nu_e}>$, $<E_{\nu_x}>$) = (11.0, 15.8) and (10.1, 12.6) MeV, respectively.
For each set of averaged neutrino energies, the event number for the Fermi distribution is between that for $\alpha=2$ and 3 of the mMB spectra, because the Fermi distribution is softer than $\alpha=2$ but harder than $\alpha=3$.
In comparison with NK1 spectra, expected event numbers of Fermi distribution with ($T_{\nu_e}$, $T_{\nu_x}$) = (3.2, 4) MeV are calculated to be smaller by about 3-5 times in spite of the larger averaged neutrino energies than NK1 (Table I).
Higher temperatures or average energies are favored to compensate for 
the missing high energy components of the spectra.
This tendency is also true for the spectra of mMB distributions 
with $\alpha\approx 3$.
Anyway, the qualitative feature that the event number becomes the largest for the MSW oscillation with the NH remains unchanged.    

In case of a failed supernova with a black hole remnant \cite{Nakazato2018}, the effects of $^{18}$O mixture are negligible as the neutrino energies are as high as $T_{\nu}$ = 6-8 MeV.    
For ($\bar{\nu}_e$, e$^{+}$) reactions, the effects can be neglected because of quite small cross sections for $^{18}$O ($\bar{\nu}_e$, e$^{+}$) $^{18}$N compared to $^{16}$O case.

\section{Summary}
Neutrino-nucleus reactions on $^{18}$O are investigated by shell-model calculations with a Hamiltonian, SFO-tls \cite{SFO}, which was used for the study of $\nu$-induced reactions on $^{16}$O \cite{SC2018}.
The GT transitions give important contributions in $^{18}$O ($\nu_e$, e$^{-}$) $^{18}$F reaction in contrary to the case of $^{16}$O, where dominant contributions come from the spin-dipole transitions.
Charged- and neutral-current reaction cross sections for $^{18}$O are evaluated by the multipole expansion method of Walecka \cite{Wal} with the use of the quenching of $g_A$ determined from the experimental GT strength obtained by ($^{3}$He, $t$) reactions \cite{ox18gt}. 

The reaction cross section for $^{18}$O ($\nu_e$, e$^{-}$) $^{18}$F is found to be larger than that for $^{16}$O due to the lower threshold energy (1.66 MeV) than that for $^{16}$O (15.42 MeV), and it remains true for low-energy neutrinos, $E_{\nu}\leq$ 20 MeV, even with the consideration of 0.205$\%$ admixture of $^{18}$O in water. 
Cross sections as functions of emitted electron energies induced by reactions on natural water are investigated at $E_{\nu}$ = 10, 20 and 30 MeV as well as for the decay-at-rest (DAR) $\nu_e$'s. 
Events from reactions on $^{16}$O and $^{18}$O are found to take place at different electron energy regions separated by 10-15 MeV. 
We have thus shown how we can separate the contributions from $^{16}$O and $^{18}$O by the measurements using the DAR $\nu_e$'s and test the calculated cross sections 
in an experiment in the near future \cite{COHERENT}.

Possible effects of the $^{18}$O admixture in water Cherenkov detectors on the evaluation of the event rate of supernova neutrinos are examined for both the case with and without the neutrino oscillations. 
Detection of events from a normal supernova at 10 kpc away 
for 32 kton water of Superkamiokande is assumed.

For neutrino spectra, NK1 \cite{Nakazato2018}, obtained by simulations of supernova explosions, the effects of $^{18}$O admixture on the event numbers are found to be modest, 6-17$\%$ increase of the count numbers.
The high energy component of the NK1 spectra, produced in the accretion phase of supernova explosions \cite{Naka2013}, suppresses relative importance of the $^{18}$O mixture in water.  
We have also shown that the effects of the neutrino oscillations are large and enhance the event numbers for the reactions on $^{16}$O and $^{18}$O by 3-4 times, which become comparable to those of the elastic $\nu$-$e^{-}$ scattering.
While these points were noticed by the previous work on $^{16}$O \cite{Nakazato2018}, 
we have confirmed these features of the cross sections and the event rate
with the inclusion of $^{18}$O and with the use of more accurate quenching factors ($q$=0.68) for the spin-dipole transitions to
the ground states of $^{16}$F 
\cite{PTEP2022}.

For neutrino spectra of modified Maxwell-Boltzmann (mMB) and Fermi distributions  
with total neutrino energy of 
3.2$\times 10^{52}$ erg/flavor,
estimated event numbers are found to depend sensitively on 
average energies 
and pinching parameters, $\alpha$,
and they increase by 1.2-3.0 (1.05-1.6) times in case without (with) the neutrino oscillations with the $^{18}$O admixture.
In any case, the change of the count numbers are similar but a bit larger than the case of the NK1 spectra; by 6-20 counts. 
The neutrino oscillations enhance the event numbers considerably similar to the case of the NK1 spectra.
Choice of neutrino spectra is important for a quantitative evaluation of the event numbers. 

We have shown for the NK1 spectra, which was obtained from an ordinary supernova neutrino model consistent with SN1987A, 
that the contribution from $^{18}$O (0.205$\%$ admixture) enhances the event rates from $^{16}$O in the electron spectra below $T_e$ = 20 MeV by 60$\%$ and 20-30$\%$ for the case without and with the neutrino oscillations, respectively.
In the latter case, the event rates for $^{18}$O and $^{16}$O become comparable to those of the neutrino-electron scattering,
while the rates below $T_e$ =20 MeV are much smaller than those of the neutrino-electron scattering, which is important to detect the direction of the supernova.
\\
\\ 

\noindent {\bf Acknowledgements}
\\
This work was supported in part by the JSPS KAKENHI, Grant Nos. JP19K03855, JP19H05811, JP20K03973 and JP20K03989.
Some part of the results in this paper was presented at the Workshop on “Neutrino Interaction Measurements for Supernova Neutrino Detection” at Oak Ridge National Laboratory.
We would like to thank the Oak Ridge National Laboratory and the workshop
participants for stimulating discussion.





\end{document}